\documentclass[intlimits,twoside,a4paper]{article}
\usepackage[cp1251]{inputenc}
\usepackage[eqsecnum]{cmpj3}
\usepackage{bm}

\issue{2021}{24}{2}{23601}
\doinumber{10.5488/CMP.24.23601}

\title{Adsorption of metallic ions from aqueous solution on
  surfactant aggregates: a molecular dynamics study}

\author [E.~H. Chavez-Martinez, E. Cedillo-Cruz,
H. Dominguez] {E.~H. Chavez-Martinez, E. Cedillo-Cruz,
  H. Dominguez\orcid{0000-0001-6126-9300} \footnote{Email address: hectordc@unam.mx}}
\address{
 Instituto de Investigaciones en Materiales, 
Universidad Nacional Aut\'onoma de M\'exico, UNAM Cd. Mx. 04510, M\'exico
}

\date{Received February 19, 2021, in final form May 21, 2021}

\begin{document}
\maketitle

\begin{abstract}

Metallic ion adsorption on surfactant aggregates
were studied with Molecular dynamics simulations.
Using ionic salts, such as lead sulfate (PbSO$_4$) and
aluminum sulfate [Al$_2$(SO$_4$)$_3$], adsorption
of lead and aluminum were investigated at different salt concentrations
and different surfactant aggregates (micelles) sizes.  The micelles
were constructed with spherical shapes composed of 
sodium dodecyl sulfate (SDS) anionic surfactants.
The electrostatic interactions between the positive
ions and the negative
SDS headgroups promote capture of the metal particles on the aggregate surface.
Metal adsorption was analyzed in terms of radial density
profiles, partial pair distribution functions and adsorption isotherms.
It is showed that SDS micelles adsorb better lead than aluminum ions
regardless of the size of the aggregates and salt concentrations.

\printkeywords
\end{abstract}

\section{Introduction}

For several years metal pollution in aquatic systems has been
the subject of a few investigations due to the severe environmental problems.
In particular, heavy toxic metals such as
lead, mercury, cadmium and aluminum, among other metallic particles,
have been the subject of various studies~\cite{wang,deng,wang1,nanseu}
as they are the cause of many health problems in humans.
For example, lead can damage the
kidneys and liver, while mercury can cause lung
damage and kidney impairment. Therefore, removal of those toxic metals from
aqueous solutions has become a very important topic not only from
a scientific point of view but also for the many industrial applications. 
Nowadays, different techniques have been used to investigate
metallic ion removal, such as chemical precipitation~\cite{ku},
ion-exchange~\cite{alyuz}, adsorption~\cite{yanagi} and
membrane precipitation~\cite{landabu}.
However, due to their polar properties, surfactant molecules have been
used as an alternative to trap metallic ions from aqueous solutions~\cite{dermont, wang1, mansoor, huang}. For instance, porous carbon
has been used with anionic and cationic surfactants to increase the removal
of metal ions.

An alternative tool to investigate such complex processes
are computer simulations. For example, Hu~et~al.~\cite{32}
and Liu et al.~\cite{34} studied adsorption and desorption
phenomena on solid surfaces. In particular, they investigated
adsorption of divalent cations and dodecane desorption from 
silica surfaces in an aqueous dilution of cetyltrimethylammonium
bromide (CTAB). On the other hand, removal studies of
organic molecules with surfactants
have also been investigated using molecular dynamics
simulations~\cite{deneb, ana, minerva, hugo}.

In a previous work, the removal
of lead and mercury ions from aqueous solutions was investigated
using an anionic surfactant, sodium dodecyl
sulfate (SDS)~\cite{edith}. For those studies, lead nitrate (Pb(NO$_3$)$_2$) and
mercury chloride (HgCl$_2$) salts were used. In that paper it was observed
that surfactants improve the retention of metal ions and they work better
for mercury than for lead.
In the present paper, lead and aluminum ions are studied
using the same  SDS anionic surfactant, though
in this case lead sulfate, PbSO$_4$ and aluminum sulfate, Al$_2$(SO$_4$)$_3$,
are tested. The use of different salts will allow us
to study the effects of salt counterions (sulfate) on the adsorption process.

\section{Computational model}

Simulations were conducted for two metal ions,
lead (Pb) and aluminum (Al) in 
lead sulfate, PbSO$_4$ and in aluminum sulfate, Al$_2$(SO$_4$)$_3$
salts, respectively. Surfactant micelles were prepared with
sodium dodecyl sulfate (SDS) molecules using
an united atom model. Each SDS molecule consisted of 
a hydrocarbon chain of 12 united carbon atoms
attached to a headgroup, SO$_4 ^-$, i.e., CH$_n$ groups were treated
as a single site atom whereas the headgroup atoms were explicitly
modelled. The neutrality of the systems
was maintained by including a Na$^{+}$ ion for each SDS molecule.

The SDS force field considered intra and intermolecular contributions
and the parameters were taken from reference~\cite{marlene},
whereas the force field for the metallic ions and for their counterions 
was taken from~\cite{maria,ljion}. It is worth mentioning that
the $\sigma$ Lennard-Jones parameters for the SO$_4$ salt
were scaled by a factor of 0.085 to better reproduce the PbSO$_4$ density.
For the SO$_4$ ion in the salt, the charges were set to  $q_{\rm{S}} = 0.288$
and $q_{\rm{O}} = -0.572$ whereas for the metallic ions, the charges
were set to $q_{\rm{Pb}} = 2.0$ for lead and
$q_{\rm{Al}} = 3.0$ for aluminum. For the aqueous media, water
molecules were used with the three-site SPC/E model~\cite{spce}.
The interactions between unlike atoms were obtained using
the Lorentz-Berthelot~(L-B) combination rules.

Initial configuration started with a single spherical micelle, previously
constructed, with  SDS molecules and placed in the center of a
cubic box. Then, the simulation box was filled with water and the micelle
was free to move. 

\begin{figure}[!t]
	\begin{center}
		\includegraphics[width=5in]{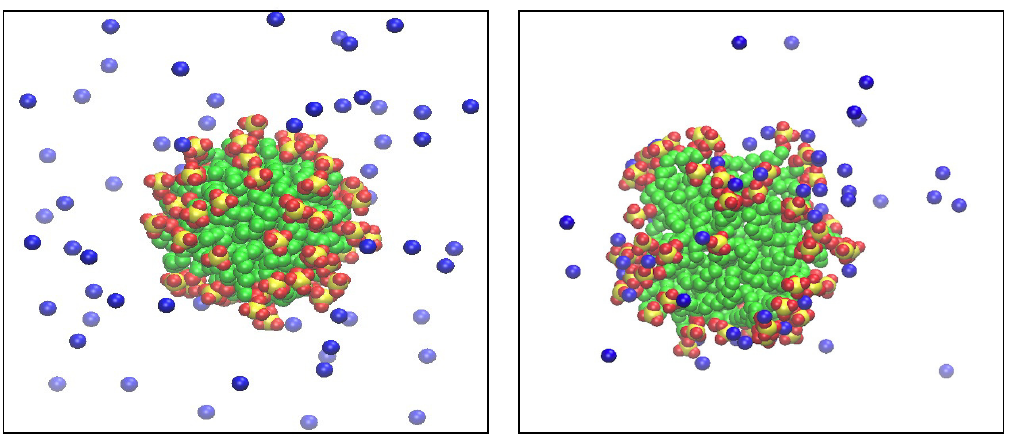}
		\caption{(Colour online) Snapshots of a SDS micelle with PbSO$_4$ salt.
			Left-hand: Initial configuration. Right-hand: Final configuration.
			Red and yellow colors
			represent the  SDS headgroups, green represents the SDS tail groups and
			blue represents the lead ions. For visualization,
			water and SO$_4$ groups are removed.}
	\label{fig-smp1}
	\end{center}
\end{figure}

Two systems were prepared with
60 and 90 SDS molecules, respectively for each system, in 11226
water molecules and they were all together equilibrated up
to 10~ns. Then, a different number of salt molecules,
PbSO$_4$ or Al$_2$(SO$_4$)$_3$, were 
randomly located in the simulation box (using individual ions
of Pb, Al and SO$_4$), i.e., 30, 60 and 90 salt molecules
completely dissolved in water, see figure~\ref{fig-smp1}.

All simulations were run in GROMACS-5.1.2
software~\cite{grom} in the NPT ensemble at temperature $T = 298.15$~K
and pressure $P = 1$~bar, using the Nos\'e-Hoover thermostat~\cite{hoover}
with a relaxation time of $\tau_T = 0.1$~ps, and the
Parinello-Rahman barostat~\cite{parinello}, with a relation time
of $\tau_p = 2$~ps, respectively.
Periodic boundary conditions were used in all directions and
long-range electrostatic interactions were
handled using the particle mesh Ewald method~\cite{essmann}.
Bond lengths were constrained using the Lincs
algorithm~\cite{lincs} and the short range interactions
were cut off at 2.0~nm. Then, simulations were carried out up to 50~ns
after 5~ns of equilibration with a timestep of $d t = 0.002$~ps. Results
were analyzed for the last 20~ns and
configuration energy was monitored as function of time to determine
when the systems reached equilibrium.
\newpage

\section{Results}

\subsection{Micelle and ions structure}

As stated above, initially a spherical micelle was placed in the center of the
simulation box. Therefore, to study whether the micelle
presents some modification in its shape
once salts were added to the system,
radii of gyration were investigated. Then, the radii of the micelles
were calculated with and without the presence of salt.
The radius of the micelle ($R_s$) is related to the radius of
gyration ($R_g$) as,

\begin{equation}
 R_s = \sqrt{\frac{5}{3}} R_g.
\end{equation}

The eccentricity of the micelle was also calculated as,

\begin{equation}
 e = 1 - {\frac{I_{\rm{min}}}{I_{\rm{avg}}}} ,
\end{equation}
where $I_{\rm{min}}$ and $I_{\rm{avg}}$ are the moment of inertia with the
minimum magnitude
and the average of all three moments of inertia, respectively.
For a sphere, this value should be zero. The radii of the micelles
without salt were 2.06~nm and 2.36~nm for the systems with 60 and 90
SDS molecules, respectively, in agreement with previous computational and
experimental results~\cite{berk,alm}. The eccentricities were 0.12 and 0.15
for micelles with 60 and 90 SDS, respectively.
When salt was included, those  values did not change significantly.
In the case of PbSO$_4$ with 60 SDS, the radii were between
2.03--2.13~nm, for
systems with 30--120~ion salts. The average eccentricity
for those micelles was $\approx 0.13$.
For the same system with 90 SDS, the radii of the micelles were
between 2.30--2.32~nm, for
systems with 30--120~ion salts. Here,
the average eccentricity of the micelles was $\approx 0.12$.
In the case of the Al$_2$(SO$_4$)$_3$ with 60 SDS, the values were similar,
for the different ion concentrations
the radii were between 2.01--2.10~nm with an average
eccentricity $\approx 0.1$.
For the system with Al$_2$(SO$_4$)$_3$ and 90 SDS, the radii,
for the different salt concentrations, were
between 2.33--2.38~nm with an average eccentricity $\approx 0.11$.
Then, it was observed that once salt was added
in the systems, the size of the micelles did not change
and since the eccentricities were small,
they remained nearly spherical, see figure~\ref{fig-smp1}.

\begin{figure}[!t]
	\begin{center}
		\includegraphics[width=2.5in]{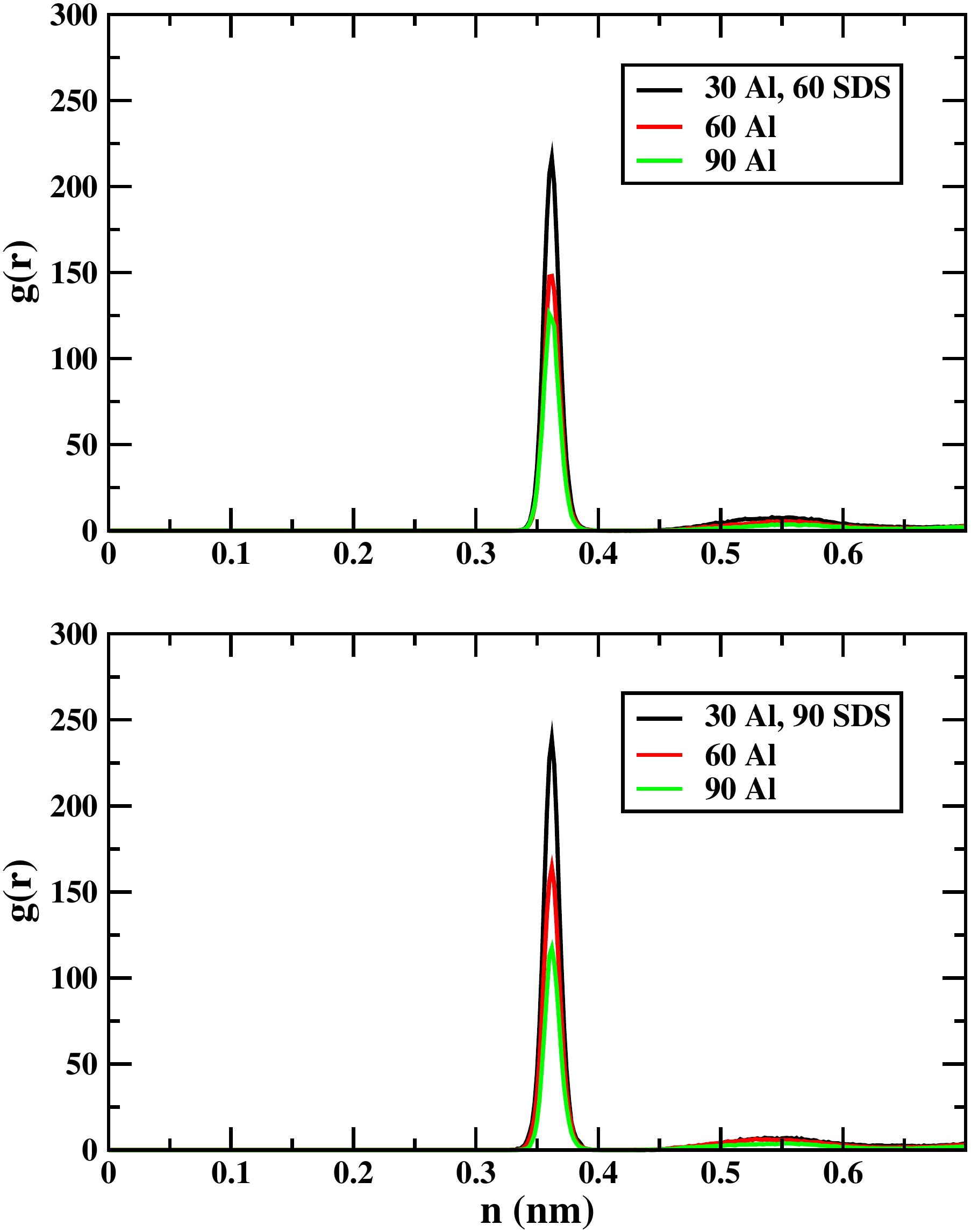}
		\caption{(Colour online) Radial distribution functions, $g_{\rm{S-Al}}(r)$, for the
			SDS micelle with different number of Al$_2$(SO$_4$)$_3$ salt ions. Top:
			with 60 SDS molecules, Bottom: with 90 SDS molecules.
			The $g_{\rm{S-Al}}(r)$ is calculated for the aluminum [Al$_2$(SO$_4$)$_3$] and
			Sulfur (SDS) pairs.}
	\label{fig-smp2}
	\end{center}
\end{figure}

At the beginning of the simulation
the ions were distributed throughout the box, at the end
some of them were adsorbed on the micelle and they were
deposited close to the SDS headgroups as it is observed in figure~\ref{fig-smp1}. In terms of partial pair correlation functions [radial
distribution functions, $g_{ij}(r)$] it was studied how the metal ions were distributed around the SDS micelles.
In figure~\ref{fig-smp2} typical radial distribution functions of the Al
ions [of the Al$_2$(SO$_4$)$_3$] with the sulfur atoms
(of the SDS surfactants) are shown. In the figure, the $g_{\rm{S}-\rm{Al}}(r)$ of the
systems with micelles of 60 SDS and 90 SDS at different Al$_2$(SO$_4$)$_3$ salt
concentrations are depicted. There is observed a first high peak suggesting
a strong interaction between the Al ions with the sulfur
atoms (SDS headgroups). It is also noted that the
peak decreases as the salt concentration increases (the number of ions),
i.e., there is less probability to find 
Al ions close to SDS headgroups at high salt concentrations.
Furthermore, higher $g_{\rm{S}-\rm{Al}}(r)$ peaks are detected
in the big micelle (90 SDS) than in the small
one at the same salt concentrations, i.e., there is less probability to find
Al ions per SDS headgroups in the small micelle. Similar
issues were observed for micelles with PbSO$_4$ salt.

\subsection{Metal ion retention}

Retention of the metal ions by the  SDS headgroups were studied in terms of
radial number density profiles,

\begin{equation}
  \rho(r) = \frac{\rd N_i(r)}{\rd V}, 
\end{equation}
$\rd N_i(r)$ is the number of ions in a spherical shell of volume
$\rd V$ ($= 4 \piup r^2 \rd r $).

\begin{figure}[!t]
	\begin{center}
		\includegraphics[width=3.4in]{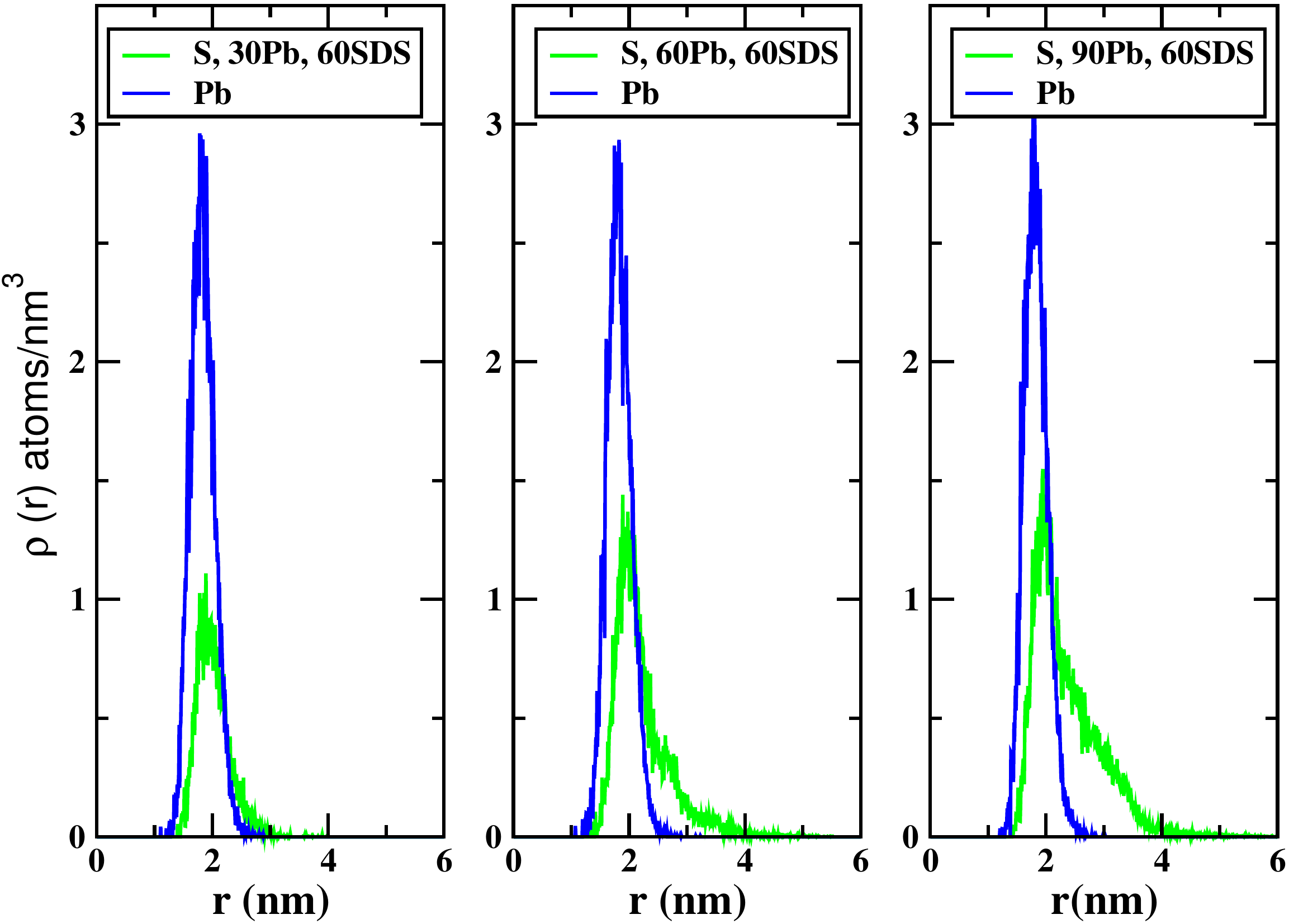}
		\caption{(Colour online) Radial density profiles for
			sulfur (SDS) and lead (PbSO$_4$) at different salt concentrations,
			i.e., the number of Pb ions, for a micelle of 60 SDS.}
		\label{fig-smp3}
	\end{center}
\end{figure}

\begin{figure}[!t]
	\begin{center}
		\includegraphics[width=3.5in]{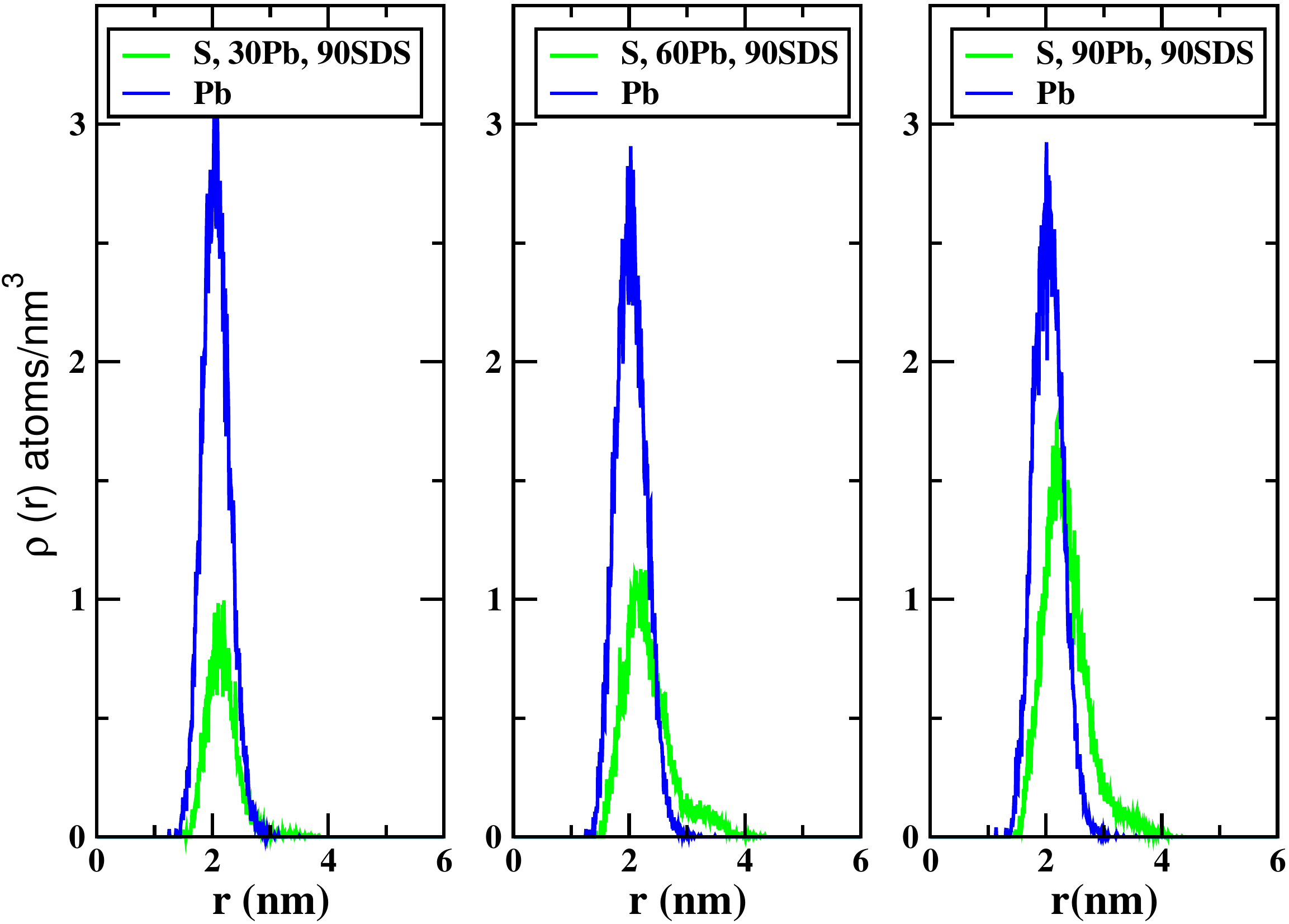}
		\caption{(Colour online) Radial density profiles for
			sulfur (SDS) and lead (PbSO$_4$) at different salt concentrations,
			i.e., the number of Pb ions, for a micelle of 90 SDS.}
		\label{fig-smp4}
	\end{center}
\end{figure}

In figures~\ref{fig-smp3} and ~\ref{fig-smp4}, typical radial density profiles
for the SDS headgroups (represented by the sulfur atoms) and Pb
metal ions are shown.
At the end of the simulations, the average
density profiles of the metal ions show a first high peak close
to the SDS peak, suggesting a few number of metal ions approaching and
adsorbed by the SDS micelle, in particular to the headgroups.
In the same plots there is also noted a shoulder and even
a tail in the ion profiles
indicating that few metals are away from the micelle, i.e., they are not held
by the SDS aggregate.

The number of ions attached on the micelle surface was calculated
by integration of the number density

\begin{equation}
  N_{ad} = \int_{0}^{R_c + r_c} 4 \piup r^2 \rho(r) \rd r.
\end{equation}

\begin{figure}[!b]
	\begin{center}
		\includegraphics[width=2.7in]{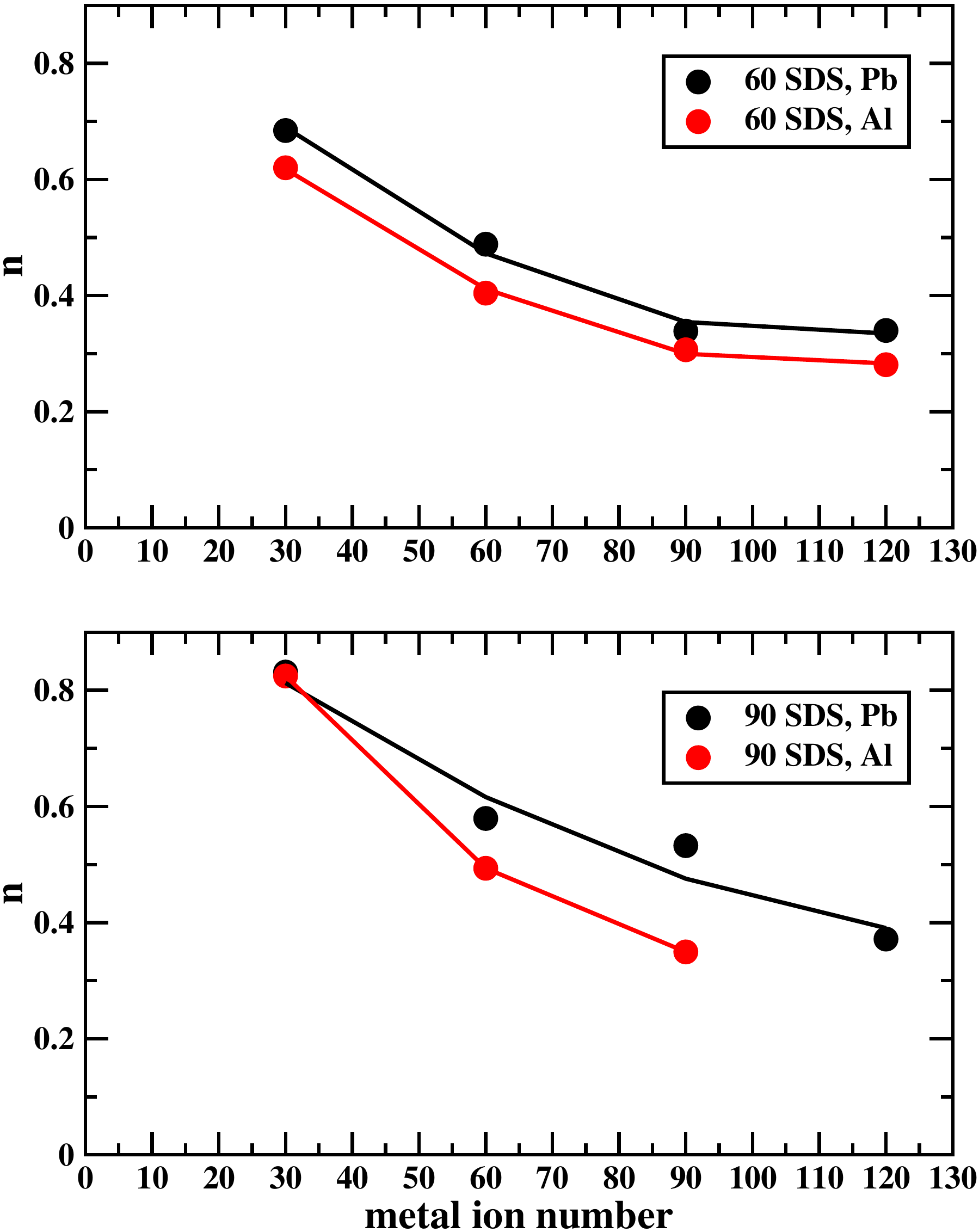}
		\caption{(Colour online) Retention ratio of metal ions (Pb or Al) at different
			salt concentrations in both micelle, 60 (top) and 90 (bottom)
			SDS. Solid lines are only
			to guide.}
		\label{fig-smp5}
	\end{center}
\end{figure}
The integration was performed in radial shells
from the center of the micelle up to the
upper limit defined by $R_c + r_c$,
where $R_c$ is the position of the SDS headgroups peak
in the density profile (see figures~\ref{fig-smp3} and~\ref{fig-smp4})  and $r_c$ is obtained by the
position of the first minimum in the radial distribution function
of the sulfur (S) with lead (Pb) [or aluminum (Al)] ions,
i.e., the position of the first
nearest neighbors of the metal ions with the S-atoms
since at this distance it is considered that the ions
are retained by the SDS micelle. In fact, the position of the sulfur peaks
of the density profiles 
can give us information on the radius of the micelle, around 2~nm,
in agreement with our previous data and with the
values of earlier simulations and
experiments reported in the literature~\cite{berk}.

Then, the amount of metallic ions retained by the SDS micelles
was estimated with the ratio $n = (N_{ad}/N_T $) calculated with
the number of metal ions adsorbed
on the SDS micelles divided by their total number ($N_T$).
The above procedure was conducted and analyzed
at different time steps throughout the entire
simulation and it was observed that after 30~ns, the number of ion
retained did not change significantly. Therefore, the results were taken
up to 50~ns, when plots reached a plateau in time.

In figure~\ref{fig-smp5}, the percentage of Al or Pb ions
retained by the different SDS micelles are showed.
As a general trend, the ion retention decreases as the salt concentration increases.
It is also observed that the big micelle retains more metallic ions
than the small one regardless the salt concentration. In fact, it is noted that
the SDS micelles work better to capture Pb ions than to capture Al ions, i.e.,
micelles adsorb more lead than aluminum at the same salt
concentration.

\subsection{Adsorption isotherms}

\begin{figure}[!t]
	\begin{center}
		\includegraphics[width=2.5in]{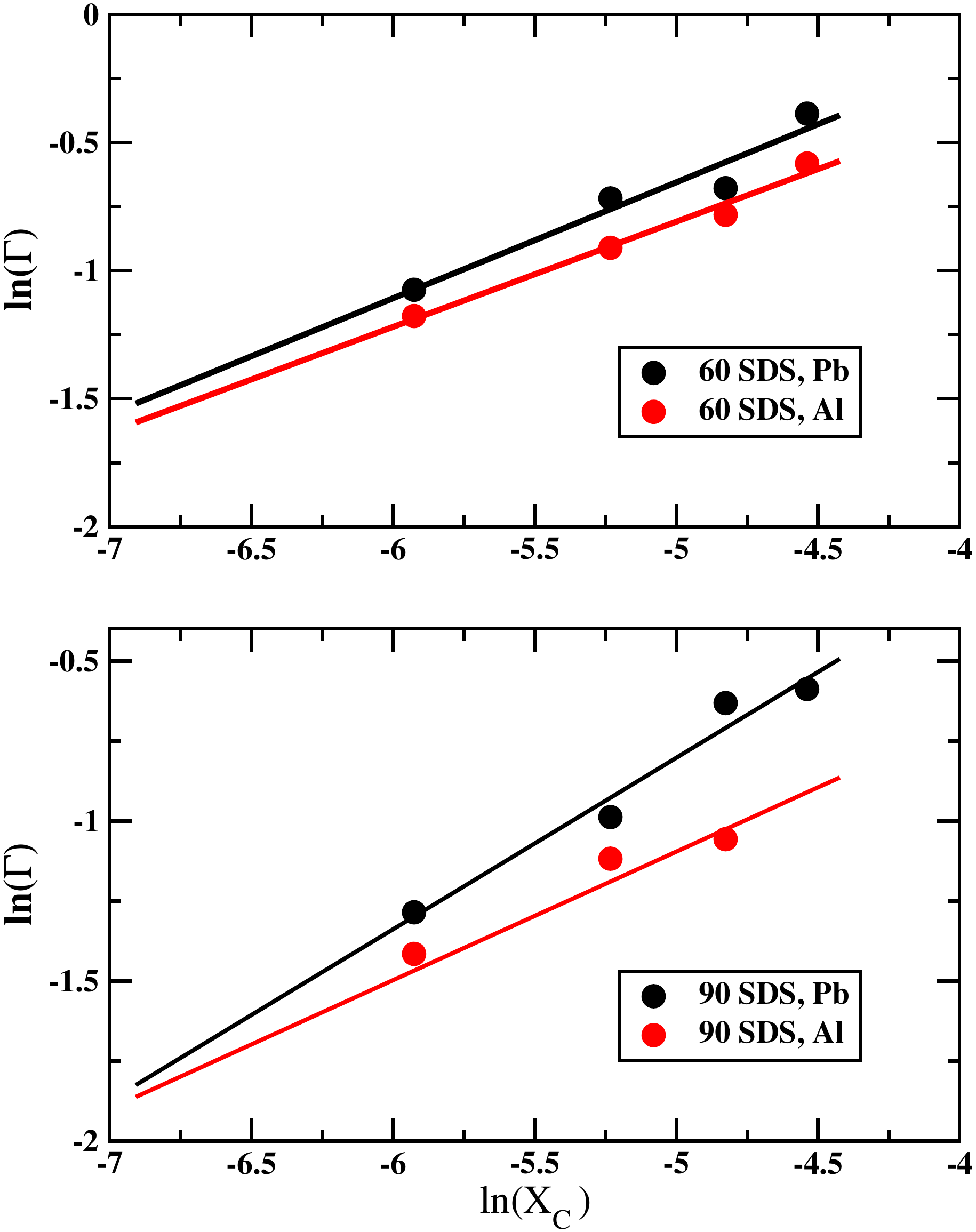}
		\caption{(Colour online) Adsorption isotherms of metal ions (Pb or Al) on SDS micelles
			at different salt concentrations. Top: micelle with 60 SDS molecules,
			bottom: micelle with 90 SDS molecules.}
		\label{fig-smp6}
	\end{center}
\end{figure}

The study can also be analyzed in terms of ion adsorption isotherms
on the SDS micelle surface. Then, adsorption was calculated
as the amount of ions attached to the micelle normalized with the
number of SDS molecules, i.e.,

\begin{equation}
\Gamma = \frac{N_{ad}}{N_{\rm{SDS}}},
\end{equation}
where $N_{ad}$ is the number of metal ions retained on the micelle
and $N_{\rm{SDS}}$ is the total number of SDS molecules in the system, i.e., the
relation between the adsorbate adsorbed and the adsorbent.
In figure~\ref{fig-smp6} adsorption isotherms are plotted, in a log-log plot,  as a
function of the ion concentration, $X_c$,

\begin{equation}
X_c = \frac{N_{\rm{ion}}}{N_{\rm{water}}},
\end{equation}
where $N_{\rm{ion}}$ and $N_{\rm{water}}$ are the total number of metal ions
and the total number of water molecules in the system, respectively.
The adsorption data of figure~\ref{fig-smp6} can be fitted with a straight line

\begin{equation}
\ln(\Gamma) = m \ln(X_c) + \ln(K),
\end{equation}
where the slope, $m$, and the intercept of the line, $\ln(K)$, can
be related to the constants of the Freundlich isotherm~\cite{isote}

\begin{equation}
\Gamma = K X_c^{1/n}.
\end{equation}
$X_c$ is the concentration, $K$ is a constant related with
the adsorption capacity and $1/n (= m)$ is the saturation rate of adsorption.
In table~\ref{tbl-smp1}, the values of $K$ and $n$ for the different systems
are given. It is observed that $K$ is higher for the systems with
Pb ions regardless of the size of the micelle, i.e.,
micelles with Pb salts have a higher capacity of adsorption
than micelles with Al salts.

\begin{table}[h]
\caption{\small{Adsorption $K$ and $n$} constants.}	 
	\label{tbl-smp1} 
\begin{center}
\begin{tabular}{|c|c|c|}
  \hline\strut
  60 SDS & $K$ & $n$ \\
  \hline
Pb & 4.99 & 2.21 \\
Al & 3.47 & 2.43 \\
  \hline
  90 SDS & $K$ & $n$ \\
    \hline
Pb & 6.53 & 1.87 \\
Al & 2.49 & 2.50 \\
\hline
\end{tabular}
\end{center}
\end{table}

\section{Conclusions}

The adsorption of lead and aluminum ions from aqueous solutions using micelles of
sodium dodecyl sulfate (SDS) surfactants, was investigated. The studies were
carried out using PbSO$_4$ and Al$_2$(SO$_4$)$_3$ salts at different
concentrations and sizes of the SDS micelles.
From the density profiles it is
observed that the negative SDS headgroups 
are close to the  positive metallic ions, and since those SDS headgroups
are located on the exterior of the micelle it can be assumed that ions
are trapped on the surface.
The partial pair distribution functions show a higher probability to find
metallic ions close to the SDS headgroups in the big micelle than
in the small one.
The results also show that SDS micelles are more efficient to retain
Pb than Al ions regardless of the aggregate size.
In fact, as a general trend, a big micelle promotes more adsorption
of metal ions than a small one, i.e.,
since big micelles have larger surfaces than small ones
there is a high probability to hold metal ions in those systems.
Moreover, the results also show that low salt concentrations
work better to hold ions on the surfactant surface.
The present work can be used to study the retention of contaminant particles
in aqueous solutions, i.e., it
shows how surfactant micelles can help trap the metal pollutant ions in water
and is complementary to the previous investigation where
SDS micelles were also used to study the
retention of lead and mercury ions~\cite{edith}.
It is worth mentioning that the results are given for a particular
combination rule between unlike atoms. With these parameters,
simulations of metallic ions-water~\cite{ljion}
and SDS-water~\cite{marlene}
have been tested in the literature with good results.
However, to the best of our knowledge, there are no reported works
on metallic ions and SDS surfactants. Different cross-interactions could give
different data, the quantitative results
may change, though we think that the qualitative results may not change,
i.e., the phenomenon will show the same trends.
Finally, from the present work and earlier results~\cite{edith} we conclude
that the counterion, in this case SO$_4$ or NO$_3$,
influences the ion retention. In fact, current data comparisons
with those of reference~\cite{edith} show that the capture of Pb ions
is better in PbSO$_4$ salts than in Pb(NO$_3$)$_2$ salts.

\section{Acknowledgements}

This work was supported by DGAPA-UNAM-Mexico grant IN105120,
Conacyt-Mexico grant A1-S-29587 and DGTIC-UNAM LANCAD-UNAM-DGTIC-238
for supercomputer facilities.
We also acknowledge Alberto Lopez-Vivas, Cain Gonzales-Sanchez and
Alejandro-Pompa for technical support in the computer calculations.

\newpage
\ukrainianpart

\title{Адсорбція іонів металів з водних розчинів на агрегатах поверхнево-активних речовин: дослідження методом молекулярної динаміки }

\author [Е. Г. Чавес-Мартінес, Е. Седільо-Крус, Е. Домінгес] {Е. Г. Чавес-Мартінес, Е. Седільо-Крус, Е. Домінгес}
\address{
Інститут матеріалознавства, Національний незалежний університет Мехіко, 04510, Mехіко, Мексика
}

\makeukrtitle

\begin{abstract}
	На основі методу молекулярної динаміки досліджується адсорбція іонів металів на агрегатах поверхнево-активних речовин.
	З використанням іонних солей, таких як сульфат свинцю (PbSO$_4$) і
	сульфат алюмінію [Al$_2$(SO$_4$)$_3$], досліджено адсорбцію
	свинцю та алюмінію для різних концентрацій солі та різних розмірів агрегатів поверхнево-активної речовини (міцел). Міцели
	будувалися у вигляді сфер, утворених аніонною поверхнево-активною речовиною -- додецилсульфатом натрію (SDS).
	Електростатичні взаємодії між позитивними іонами і від'ємно зарядженими групами
	SDS сприяють захопленню частинок металу поверхнею агрегата.
	Адсорбцію металів досліджено на основі аналізу профілів радіальної густини, парціальної парної функції розподілу та ізотерм адсорбції.
	Показано, що міцели SDS краще адсорбують іони свинцю, ніж алюмінію,
	незалежно від розміру агрегатів чи концентрації солі.
	\keywords адсорбція іонів металів, аніонна поверхнево-активна речовина,
	поверхнева адсорбція, молекулярна динаміка, іонні солі
\end{abstract}


\begin{thebibliography}{1}

\bibitem{wang} Wang H., Zhou A., Peng F., Yu H., Yang J., 
  J. Colloid Interface Sci., 2007, \textbf{316}, 277,\\
  \doi{10.1016/j.jcis.2007.07.075}.

\bibitem{deng} Deng L. P., Su Y. Y., Su H., Wang X. T., Zhu X. B.,
  J. Hazard. Mater., 2007, \textbf{143}, 220,\\ \doi{10.1016/j.jhazmat.2006.09.009}.

\bibitem{wang1} Wang J., Feng X., Andersen C. W. N., Xing Y., Shang l.,
  J. Hazard. Mater., 2012, \textbf{221}, 1, \\
  \doi{10.1016/j.jhazmat.2012.04.035}.

\bibitem{nanseu} Nanseu-Njiki C. P., Tchamango S. R., Ngom P. C.,
  Darchen A., Ngameni E., J. Hazard. Mater., 2009, \\\textbf{168}, 1430,
  \doi{10.1016/j.jhazmat.2009.03.042}.

\bibitem{ku} Ku Y., Jung I. L.,  Water Res., 2001, \textbf{35}, 135,
  \doi{10.1016/S0043-1354(00)00098-1}.

\bibitem{alyuz} Alyuz B., Veli S.,
  J. Hazard. Mater., 2009, \textbf{167}, 482,
  \doi{10.1016/j.jhazmat.2009.01.006}.

\bibitem{yanagi} Yanagisawa H., Matsumoto Y., Machida M.,
 Appl. Surf. Sci., 2010, \textbf{256}, 1619,
 \doi{10.1016/j.apsusc.2009.10.010}.
 
\bibitem{landabu} Landaburu-Aguirre J., Garc\'ia V.,
  Pongr\'acz E., Keiski R. L., Desalination, 2009, \textbf{240}, 262,\\
  \doi{10.1016/j.desal.2007.11.077}

\bibitem{dermont} Dermont G., Bergeron M., Mercier G., Richer-Lafleche M.,
  J. Hazard. Mater., 2008, \textbf{152}, 1,\\
  \doi{10.1016/j.jhazmat.2007.10.043}.

\bibitem{mansoor} Anbia M., Amirmahmoodi S.,
  Arabian J. Chem., 2016, \textbf{9}, S319,
\doi{10.1016/j.arabjc.2011.04.004}.

\bibitem{huang} Huang C. P., Blankenship D. W.,
  Water Res., 1984, \textbf{18}, 37,
  \doi{10.1016/0043-1354(84)90045-9}.

\bibitem{32} Hu X., Li Y., Sun H., Song X., Li Q., Cao X., Li Z.,
  J. Phys. Chem. B, 2010, \textbf{114}, 8910,
  \doi{10.1021/jp101943m}.

\bibitem{34} Liu Q., Yuan S., Yan H., Zhao X.,
  J. Phys. Chem. B, 2012, \textbf{116}, 2867,
  \doi{10.1021/jp2118482}.

\bibitem{deneb} Peredo-Mancilla D., Domimguez H., J. Mol. Graphics Modell., 2016,
  \textbf{65}, 108,
  \doi{10.1016/j.jmgm.2016.02.011}.

\bibitem{ana} Salazar-Arriaga A. B., Domimguez H.,
  Chem. Phys., 2020, \textbf{539}, 110945,
 \doi{10.1016/j.chemphys.2020.110945}.
  
\bibitem{minerva} Valencia-Ortega M., Fuentes-Azcatl R., Dominguez H.,
  J. Mol. Graphics Modell., 2019, \textbf{92}, 243,\\
  \doi{10.1016/j.jmgm.2019.08.003}.
  
\bibitem{hugo} Espinosa-Jimenez H., Dominguez H.,
  Rev. Mex. Fis., 2019, \textbf{65}, 20.

\bibitem{edith} Cedillo-Cruz E., Garcia-Ramos D., Dominguez H.,
  Chem. Phys. Lett., 2021, \textbf{767}, 138340,\\
  \doi{10.1016/j.cplett.2021.138340}.

\bibitem{marlene} Rios-Lopez M., Mendez-Bermudez J. G., Dominguez H.,
  J. Phys. Chem. B, 2018, \textbf{122}, 4558,\\
  \doi{10.1021/acs.jpcb.8b01452}.

\bibitem{maria} Pacheco-Blas M. del A., Dominguez H., Rivera M.,
  Chem. Phys., 2017, \textbf{485}, 13,\\
  \doi{10.1016/j.chemphys.2017.01.002}.

\bibitem{ljion} Heinz H., Vaia R. A., Farmer B. L., Naik R. R.,
  J. Phys. Chem. C, 2008, \textbf{112}, 17281,
\doi{10.1021/jp801931d}.

\bibitem{spce} Berendsen H. J. C., Grigera J. R., Straatsma T. P.,
  J. Phys. Chem., 1987, \textbf{91}, 6269,
  \doi{10.1021/j100308a038}.

\bibitem{grom} Hess B., Kutzner C., van der Spoel D., Lindahl E.,
  J. Chem. Theory Comput., 2008, \textbf{4}, 435,\\
  \doi{10.1021/ct700301q}.

\bibitem{hoover} Hoover W. G., Phys. Rev. A, 1985, \textbf{31}, 1695,
  \doi{10.1103/PhysRevA.31.1695}.
  
\bibitem{parinello} Parrinello M., Rahman A.,
  J. Appl. Phys., 1981, \textbf{52}, 7182,
  \doi{10.1063/1.328693}.

\bibitem{essmann} Essmann U., Perera P., Berkowitz M. L.,
  Darden T., Lee H., Pedersen L. G., J. Chem. Phys., \\ 1995,  \textbf{103}, 8577,
  \doi{10.1063/1.470117}.
  
\bibitem{lincs} Hess B., Bekker H., Berendsen H. J. C., Fraaije J. G. E. M.,
  J. Comput. Chem., 1997, \textbf{18}, 1463,\\ \doi{10.1002/(SICI)1096-987X(199709)18:12<1463::AID-JCC4>3.0.CO;2-H}.
 
\bibitem{berk} Bruce C. D., Berkowitz M. L., Perera L., Forbes M. D. E.,
  J. Phys. Chem. B, 2002, \textbf{106}, 3788,\\
  \doi{10.1021/jp013616z}.

\bibitem{alm} Almgrem M., Swarup S., J. Phys. Chem., 1982, \textbf{86}, 4212,
  \doi{10.1021/j100218a024}.

\bibitem{isote} Freundlich~H., Kapillarchemie: eine Darstellung
  der Chemie der Kolloide und verwandter Gebiete, Akademische
  Verlagsgesellschaf, Leipzig, 1909.

\end{thebibliography}
\end{document}